\documentclass[11pt,letterpaper]{cslreport}
\usepackage{graphicx}
\usepackage{amsmath}
\usepackage{rotating}
\usepackage{url}
\usepackage{alltt}
\usepackage{color}
\usepackage{xcolor}
\usepackage{boldline,multirow}
\usepackage{colortbl}
\usepackage{fancyhdr}
\usepackage{tabularx}
\usepackage{listings}
\usepackage{amssymb}
\usepackage{fancybox}
\newlength{\hsbw}
 \usepackage[inline]{enumitem}
\newlist{inlinelist}{enumerate*}{1}
\setlist*[inlinelist,1]{%
  label=(\arabic*),
  itemsep=5pt,
}

\usepackage{soul}
\usepackage{hyperref}
\hypersetup{
    colorlinks=true,
    linkcolor=blue,
    filecolor=magenta,
    urlcolor=cyan,
    pdftitle={Report on Evidential Transactions},
    pdfpagemode=FullScreen,
    }

\usepackage{wrapfig}
\definecolor{light-gray}{gray}{0.95}

\usepackage{verbatimbox}
\usepackage{fancyvrb}
\ttfamily

\usepackage[english]{babel}
\usepackage[utf8x]{inputenc}
\usepackage[T1]{fontenc}

\usepackage{amsmath}
\usepackage{amsfonts}
\usepackage{graphicx}
\usepackage{pdfpages}

\newcommand{\Land}{\,\land\,}
\newcommand{\Lor}{\,\lor\,}
\newcommand{\Imp}{\,\Rightarrow\,}
\newcommand{\Bimp}{\,\Leftrightarrow\,}
\newcommand{\Lneg}{\lnot\,}
\newcommand{\False}{\bot}
\newcommand{\True}{\top}

\newcommand{\Exists}[2]{(\exists  #1)\,#2}
\newcommand{\Forall}[2]{(\forall  #1)\,#2}

\newcommand{\Cyberlogic}{{{{\em Cyberlogic}}}}
\newcommand{\Dom}[1]{Dom(#1)}

\newcommand{\D}{D}
\newcommand{\G}{G}
\newcommand{\Atom}{Atom}

\newcommand{\Prv}[1]{Prv(#1)}
\newcommand{\Pub}[1]{Pub(#1)}
\newcommand{\Encr}[2]{Encr(#1,\, #2)}
\newcommand{\Decr}[2]{Decr(#1,\, #2)}
\newcommand{\Hash}[1]{Hash(#1)}

\newcommand{\Attest}[2]{{#1}\,\rhd\,{#2}}
\newcommand{\AttestBefore}[3]{{#1}\,\rhd_{{\small{\mathrm{bef}({#2})}}}\,{#3}}
\newcommand{\AttestAfter}[3]{{#1}\,\rhd_{{\small{\mathrm{aft}({#2})}}}\,{#3}}
\newcommand{\AttestAt}[3]{{#1}\,\rhd_{{\small{\mathrm{at}({#2})}}}\,{#3}}
\newcommand{\Delegate}[3]{\mathrm{DL}[{#1}\,\to\,{#2}] \,{#3}}

\newcommand{\AttestEventually}[2]{\Diamond({#1}\,\triangleright\,{#2})}
\newcommand{\AttestGlobalInterval}[4]{\Box_{[{#1},{#2}]}({#3}\,\triangleright\,{#4})}

\newcommand{\Knows}[2]{\mathrm{KB}_{#1}({#2})}

\title{
Evidential Transactions with Cyberlogic
}

\author{
 \textbf{Harald Ruess} \\
  fortiss \\
  Guerickestra{\ss}e 25 \\
   D-80805 M\"unchen \\
  Phone: (089) 3603522-11 \\
 \texttt{ruess@fortiss.org} \\[3mm]
\and
  \textbf{Natarajan Shankar}\\
  SRI International \\
  333 Ravenswood Avenue  \\
  Menlo Park CA 94025 \\
  Phone: (650) 859-5272 \\
  \texttt{natarajan.shankar@csl.sri.com}
 }

\date{CSL Technical Report SRI-CSL-2023-01 • 11 January, 2023 \vspace{1cm}}

\begin{document}

\rhead{\includegraphics[width=0.4\columnwidth]{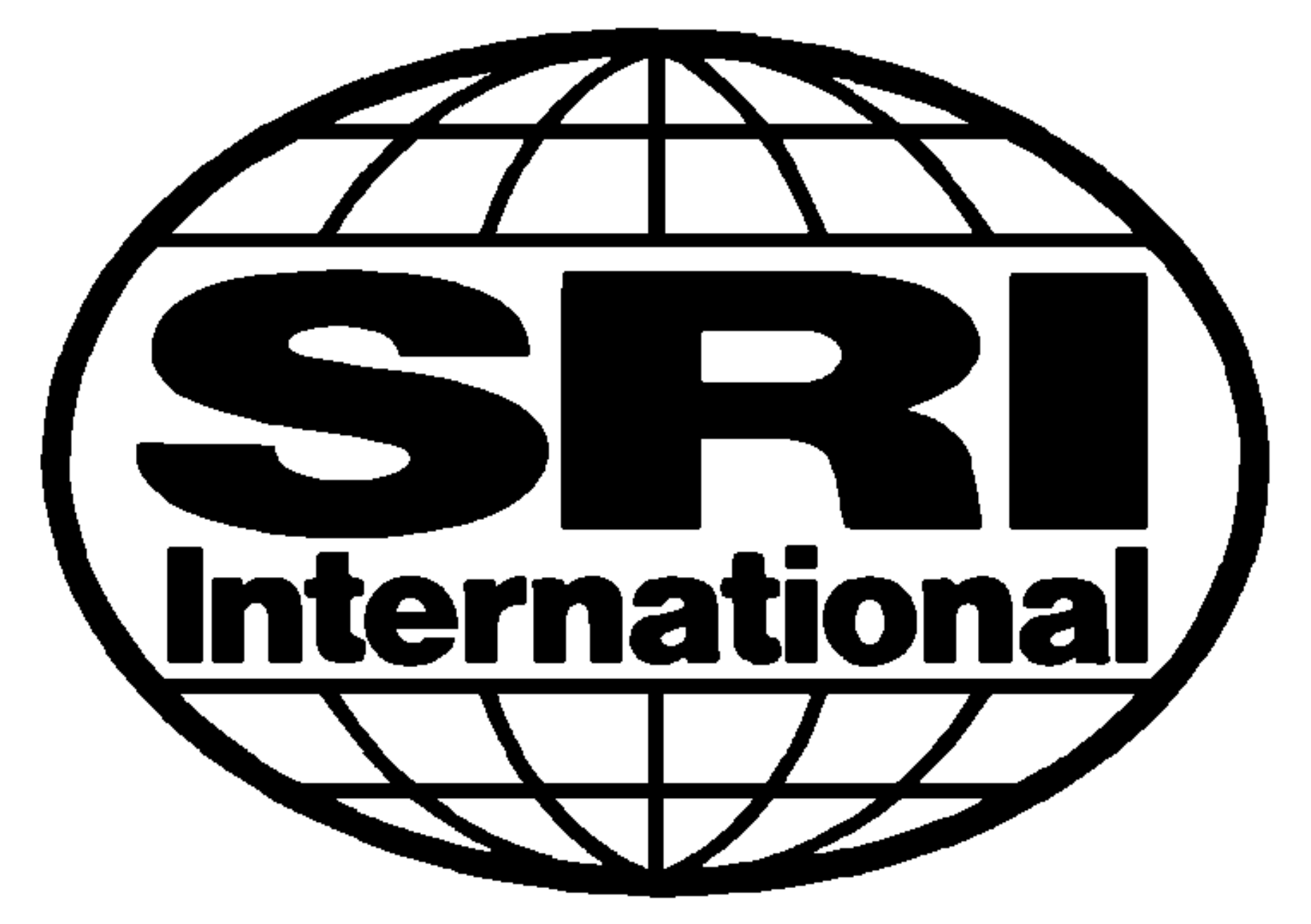}}
\renewcommand{\headrulewidth}{0.4pt}
\renewcommand{\footrulewidth}{0.4pt}

\maketitle

\vspace{3.5cm}

\noindent
{\footnotesize 
\bfseries
This research has been supported by the DARPA Automated Rapid Certification Of Software (ARCOS) project under contract number HR0011043439, National Institute of Aerospace Award \#C21-202017-SRI, and NSF Grant SHF-1817204.  Any opinions, findings, and conclusions or recommendations expressed in this material are those of the author(s) and do not necessarily reflect the views of the United States Government, DARPA, NASA, NIA, or NSF.}

%

%
%

\newpage
\pagenumbering{arabic}
\setcounter{page}{2}


\fancyhead{}
\fancyhead[l]{\footnotesize SRI International}
\fancyhead[c]{\footnotesize Cyberlogic}
\fancyhead[r]{\footnotesize January 2023}
\renewcommand{\headrulewidth}{0.4pt}
\renewcommand{\footrulewidth}{0.4pt}


%
\newpage
\tableofcontents
\setcounter{page}{0}  

\begin{abstract}
Cyberlogic is an enabling logical foundation for building and analyzing
digital transactions that involve the exchange of 
digital forms of evidence. 
It is based on an extension of (first-order) intuitionistic predicate logic with 
an attestation and a knowledge modality.
The key ideas underlying Cyberlogic are extremely simple, as (1) public keys correspond to authorizations, 
(2) transactions are specified as distributed logic programs, and
(3) verifiable evidence is collected by means of distributed proof search.
Verifiable evidence, in particular, are constructed  from extra-logical elements such as signed documents and cryptographic signatures.
Despite this conceptual simplicity of Cyberlogic,  central features of authorization policies 
including trust, delegation, and revocation of authority are definable.
An expressive temporal-epistemic logic for specifying distributed authorization policies and protocols 
is therefore definable in Cyberlogic using a trusted time source.
We describe the distributed execution of Cyberlogic programs based on the hereditary Harrop fragment in terms of distributed proof search, 
and we illustrate some fundamental issues in the distributed construction of certificates. 
The main principles of encoding and executing cryptographic protocols in Cyberlogic are demonstrated. 
Finally, a functional encryption scheme is proposed for checking certificates of evidential transactions when policies are kept private.

\end{abstract}

\begin{flushright}
~\\
\begin{small}
{\em "If electronic mail systems are \\
to replace the existing mail system for business transactions, \\
{\em signing} an electronic message must be possible. \\
The recipient of a  signed message has proof \\
that the message originated from the sender. \\
This quality is stronger than mere authentication (\ldots); \\
the recipient can convince a  {\em judge} \\
that the signer sent the message."} \\
   (Rivest, Shamir, Adleman~\cite{rivest1978method})
\end{small}
\end{flushright}

\chapter{Introduction}

We have outsourced to Internet technology all kinds of transactions and social interactions that previously required the mutual attribution of trust and responsibility.
Many of the entities involved in Internet transactions, however, are unknown to each other, and there usually is no central authority that everyone trusts.
But then, how can we trust that digital transactions are being executed "on the Internet" as intended? How can we trust that crucial information is not being misused or leaked? 
How can we ensure that we are not being deceived or otherwise being robbed?
How can we assure responsibility and accountability for actions?

As of now we do not have  convincing answers for trustworthy Internet transactions, even though they are
already woven into our very societal fabric.
%
%
Recent attempts of alleviating the situation based on voluntary self-commitment of all stakeholders~\cite{berners2019invented}\@ have not been proven themselves to be effective. 

Traditional identity-based trust management is also not applicable to Internet transactions.
In older settings of access control, in particular, authorization of a request is divided into 
authentication({\em 'who made the request?'}) and access control ({\em 'is the requester authorized to perform the action?'})\@.
But this may be ineffective when the resource owner and requester are unknown to one another.  
Because the access control mechanism does not know the requester directly, it has to use information from  third parties  
who know, or claim to know, the requester better. 
But, normally, the authorizer trusts these third parties only for certain things and only to certain degrees.
This trust and delegation aspect makes Internet transactions different from more traditional access control~\cite{li2003delegation}\@.

Distributed immutable ledgers such as blockchains allow for digital transactions with little reliance on 
trusted third-parties.
This lack of trustworthy parties in blockchains, however, has serious consequences for performance, scalability, and decentralization~\cite{chu2018curses}\@.  
In addition, blockchain-based transactions  occur in the context of complex programs, which are sometimes referred 
to as {\em  smart contracts}\@.
As such a smart contract foregoes trusted third-parties and runs on several computers anonymously, it 
constitutes a highly critical program that has to be trusted-by-design~\cite{dill2021fast} together with its underlying distributed execution (and miner) framework. 
In this way, a large number of vulnerabilities coming from the execution of the programs in a widely distributed blockchain network are fully exploitable today~\cite{atzei2017survey}\@.\footnote{
For instance, the DAO exploit \\
   ({\tt http://hackingdistributed.com/2016/06/18/ analysis-of-the-dao-exploit/})\@.
}

Here, we follow the {\em 'verify, then trust'}  paradigm of authorization in open distributed systems~\cite{grandison2000survey,blaze1996policymaker,li2003delegation}, 
where a {\em requester} submits a request, possibly supported by a set of {\em credentials} issued by other 
parties, to an {\em authorizer} who controls the requested resources.
The identity information is just one kind of credential, and it may be necessary and sufficient for some applications but not for others.
The authorizer then decides whether to authorize this request by answering the {\em proof-of-compliance} question: {\em 'do these credentials prove that a request complies with my policy?'}\@. 

Consider, for example, the process of obtaining a visa, which despite its conceptual simplicity
is often overly tedious and error-prone in real life.
The requester initiates a transaction for obtaining a visa by contacting an authorized visa-issuing consulate and by providing any number of additional credentials such as  deeds, certified translations, and affidavits as requested.
In the course of authorizing a visa request, the consulate usually contacts other authorities for additional information and background checks with, among others, banks, health insurance providers, airlines, and keepers of {\em no-fly} lists.
In case these checks are positive, the authorized consulate may decide to issue a visa, which in our case is just a digital certificate signed by the private key of the consulate. 
This certificate attests to the fact that the consulate has issued the desired visa to the requester.
Now, upon entry, the requester presents her visa certificate, and the border control agent checks, 
by decrypting the certificate with the corresponding public key of the issuing consulate and possibly some additional checks,  that the requester indeed is eligible for entering the country.
It is easy to extrapolate from the above scenario to other uses of digital evidence in electronic commerce, business and administrative processes, and digital government. 
The key observation is that these transactions usually involve the exchange of various forms of evidence.

Cyberlogic is an enabling foundation for building and analyzing digital transactions that involve the exchange of 
evidence~\cite{ruess2003introducing,bernat2006first}\@. 
It builds on the existing public key infrastructure while noting that the utility of such an infrastructure depends on a coherent and 
rigorous semantic foundation.
Without such a foundation, terms like authentication, anonymity, trust, and certification are not meaningful, and the resulting protocols 
cannot by themselves provide the necessary guarantees. 
The correct functioning of Cyberlogic is primarily based on the correct specification of Cyberlogic policies, the correct functioning of the underlying public key infrastructure and the unbreakability of its ciphers.
Therefore, and in sharp contrast to blockchains,  there is no need in Cyberlogic to trust smart contracts and their distributed execution environment, as the {\em kernel of trust} in Cyberlogic is reduced to cryptographic proof checking.

Cyberlogic allows (1) for expressing transactional policies and protocols of a distributed system's interacting actors, 
called principals in this context,
(2) for the distributed execution of transactions, and 
(3) for generating tamper-proof verifiable evidence. 
Each principal receives queries from other principles, performs computation and deduction in accordance with their policies and their current knowledge, and communicates responses and subqueries to other principals.
The policy of the visa processing actor above might require to subquery, among others, bank and insurance principals for certifying the requestee's bank account and insurance status. 
Furthermore, the border control agent's policy might permit entry only after checking that the presented visa certificate has not been revoked recently.
The visa itself is presented in the form of logical relations such as  $hasVisa("John Doe", 2022, 90)$, which are signed by the private key of an authorized consulate.
Consequently, the validity of the presented evidence can be checked by the border control agent by decrypting this certificate 
with the corresponding public key\@. 

Altogether, Cyberlogic is a logic for {\em evidential} transactions built on the public-key infrastructure. 
The key ideas underlying Cyberlogic are extremely simple.
\begin{enumerate}
    \item Private/Public key pairs correspond to authorizations;
    \item Transactional policies are specified as distributed logic programs;
    \item Verifiable evidence is constructed by distributed proof search.
\end{enumerate}
In this way, logical statements $\varphi$ are signed by private keys $\Prv{K}$ to obtain a digital certificate $c$, so that $c$ is evidence for the claim that (principal/authority) $K$ attests to $\varphi$\@. 
This is verified by decrypting $c$ with the corresponding public key $\Pub{K}$ to see if it yields $\varphi$\@. 
Notice that the fact that authority $K$ attests to some logical statement does not necessarily imply its logical validity.

Transactional policies and protocols are distributed logic programs that gather evidence by using both ordinary 
predicates and digital certificates, 
and the execution of distributed logic programs corresponds to distributed  proof search.
Execution is triggered by a query which itself may trigger, 
based on the local policies and knowledge, any number of subqueries.
The communication between principals therefore involves queries and (partial) answers to these queries. 

In the visa example above, the authorized consulate provides a visa certificate to the requester, which (or its hash) is 
signed by private key of the consulate. Upon entry of the requester, the border control agent verifies the visa certificate by decrypting the provided certificate with the corresponding public key.
Real world authorization schemes, however, are considerably more complicated. 
For instance, the authority of handing out visas might have be dynamically delegated through a chain of authorizations by, say, the state department and embassies.
Also, authorizations such as visas may be revoked due to a number of different reasons.  
This complexity and dynamicity of distributed policies
suggests declarative specifications by means of an expressive logic.

This exposition is structured as follows. 
Chapter~\ref{sec:logic} describes the logical foundation of Cyberlogic, which is based on an extension of intuitionistc 
predicate logic with an attestation modality.
Chapter~\ref{sec:proof.search} illustrates distributed proof search in Cyberlogic and the interactive, incremental construction of
certificates based on distributed proof search. 
Then, in Chapter~\ref{sec:expressiveness} we demonstrate the expressive power of Cyberlogic by encoding all-important policy mechanisms such as authorization and delegation (Section~\ref{sec:delegation}), timed attestation (Section~\ref{sec:time}), revocation of authorities (Section~\ref{sec:revocation}), and the generation of fresh, unique numbers (Section~\ref{sec:nonces})\@. 
Distributed execution for the hereditary Harrop fragment is outlined in Chapter~\ref{sec:execution} and illustrated,
by means of the infamous Needham-Schroeder authentication protocol in Chapter~\ref{sec:needham.schroeder}\@.
Communication of tree-like certificates in Cyberlogic might be avoided using the proposed notions of cryptographic proofs in Chapter~\ref{sec:crypto.proofs}\@. 
We conclude with a comparison of Cyberlogic to most-closely related work in Chapter~\ref{sec:related.work} and some final remarks in Chapter~\ref{sec:conclusions}\@. 

\paragraph{Disclaimer.}
This report should most probably have already been finished long time ago, as it is largely based on an invited talk "Introducing Cyberlogic" at the 
{\em High Confidence Software and Systems} conference in Baltimore on April 1, 2003. 
The original development of Cyberlogic  has been supported by the National Science Foundation under the grant No. 0208779, and the preparation of this report has been supported by the Bavarian Ministry of Economics in the context of the fortiss AI Center.
We are revisiting Cyberlogic here due to new and increased interest on trust management for open and large-scale distributed systems such as the European GAIA-X initiative and decentralized, automated certification of software in DARPA's ARCOS project. 
In particular, we thank Vivek Nigam for his constructive comments on a previous version of this paper, and Ulrich Sch{\"o}pp for suggesting multi-input functional encryption for ensuring the secrecy of proof checkers.

\chapter{Attestation and Knowledge}  \label{sec:logic}

Cyberlogic is an intuitionistic (first-order) predicate logic~\cite{dummett2000elements} extended with an {\em attestation} modality\@. In particular, signatures $\Sigma$ are multi-sorted, and each sort $S$ has an associated domain $\Dom{S}$ of interpretation. 
Signatures are assumed to contain a sort $Principal$ with a countably 
infinite supply of constant symbols,
and there is a countably infinite supply of logical variables for each sort.
Atomic formulas are either $\True$, $\False$, relations of the form $p(t_1, \ldots, t_n)$ with $p$ a relational symbol in $\Sigma$ and $t_i$, for $i=1,\ldots,n$, correspondingly well-sorted $\Sigma$-terms with variables. 
Logic formulas are built up recursively in the usual way, and include atomic formulas, attestations of the form $\Attest{K}{\phi}$ with $K$ a constant symbol of sort $Principal$ and $\phi$ a formula, conjunctions $\phi \Land \psi$, disjunctions $\phi\Lor \psi $, implications $\phi\Imp\psi$, universal quantification $\Forall{x:S}{\phi}$, where $x$ is a logical variable of sort $S$ in $\Sigma$, and, similarly, existential quantification $\Exists{x:S}{\phi}$\@. 
Negation $\Lneg{\phi}$ is defined as $\phi \Imp\False$ and bi-implication $\phi\Bimp\psi$ stands for $\phi\Imp\psi \Land \psi\Imp\phi$\@.  

Some symbols in $\Sigma$ may be {\em interpreted} with respect to a first-order theory\@. For example, symbols such as $+$,  $=_{Nat}$ are interpreted in the theory of arithmetic (say, Heyting or Peano)\@.
A sentence is a formula $\phi$  without any free variables, that is, all variables are bound by a universal or existantial quantification. 
Moreover, substitution in $\phi$ of a free variable $x$ for a well-sorted term $t$, which does not contain $x$, 
of the same sort is denoted by $\Phi[x/t]$\@. 

The attestation modality $\Attest{K}{\phi}$ expresses the fact that principal $K$ attests to the statement $\phi$\@.
Attestation is, of course, in the tradition of the {\em says} modality~\cite{abadi1993calculus} in access control 
logics, and its introduction rules are derived from the {\em says} modality in lax logics~\cite{fairtlough1997propositional,davies2001modal}; more precisely:
    $\phi \Imp \Attest{K}{\phi}$
    and 
    $(\phi \Imp \Attest{K}{\psi}) \,\Imp\,         
                 (\Attest{K}{\phi} \Imp \Attest{K}{\psi}) 
    $
hold for all principals $K$ and for all formulas $\phi$, $\psi$\@. 
In particular, any principal can attest to any valid sentence but not every sentence a principal attests to is valid;  that is, 
     $
     (\Attest{K}{\phi}) \not\Imp \phi
     $ holds for all formulas $\phi$\@.
Moreover, the attestation modality is absorbing ($\Attest{K}{\Attest{K}{\phi}} \,\Bimp\, \Attest{K}{\phi}$)
but not commutative ($\Attest{K}{\Attest{L}{\phi}} \not\Imp \Attest{L}{\Attest{K}{\phi}}$\@.     
   
Reconsidering the visa example, somebody named {\em "John Doe"} requests a travel visa for 14 days in 2022.
He initiates his request by issuing the query $\Attest{Cons42}{visa("John Doe", 2022, 90)}$ to the 
principal $Cons42$, which is the authorized visa-issuing consulate\@.
This consulate may choose to respond, after some further authentication and inquiries about financial status and potential police records, to this request by attesting  that  {\em "John Doe"} is eligible for entering the country in 2022 on a tourist visa with a maximum stay of 90 days. 
In this case, the  consulate {\em Cons42} responds by attesting
$\Attest{Cons42}{visa("John Doe", 2022, 90)}$\@.
This concludes the transaction between {\em "John Doe"} and the consulate {\em Cons42}\@.
Such a response, however, is only of informative nature, as anybody can make up and present these kinds of claims. 
Therefore, upon entry into the country, the border control does not only need to authenticate "John Doe" but 
it also needs to verify his visa status by querying for a corresponding entry in the data base of the responsible authorities.
In an evidential transaction, the authorizer {\em Cons42} provides verifiable evidence that the requester {\em "John Doe"} indeed is eligible for entry to the country on a tourist visa. In this way, the border control may directly check this evidence for a visa, and permit after verifying {\em "John Doe"}'s identity, for example by checking his passport, but without any further consultations to other visa-granting authorities.  

Think of this evidence as being represented as a number or a digital certificate. Also, authorities such as the visa-granting consulate {\em Cons42} have an associated public-private key pair $(\Prv{Cons42}, \Pub{Cons42})$\@. In general, every constant symbol $K$ of sort {\em Principal} has an associated key pair $(\Prv{K}, \Pub{K})$ such that\footnote{Syntactic equivalence instead of bi-implication usually suffices.}
$$
     \Decr{\Pub{K}}{\Encr{\Prv{K}}{\varphi}} \,\Bimp \, \varphi\mbox{\@.}
$$
In this way, the consulate uses its private key $\Prv{Cons42}$ to sign the relation (or its hash)
$usTouristVisa("John Doe", 2022, 90)$ and return this signature as evidence to the requestee. 
The border control then verifies the presented evidence simply by decrypting it using the public key of {\em Cons42}\@. 
In general, evidence for an attestation $\Attest{K}{\varphi}$ is a number $c$ such that $\Decr{\Pub{K}}{c}$ yields $\varphi$\@. 
In this case we also say that $c$ realizes this attestations, and we write:
     $$c \,:\, \Attest{K}{\varphi}\mbox{\@.}$$
Based on the Brouwer-Heyting-Kolmogorov (BHK) explanation of intuitionistic truth, the notion of realizability is extended in the usual way to all intuitionistic sentences.
\begin{eqnarray*}
       *  & : & \True \\
     (c,d) & : &  (\phi \Land \psi)             ~~~~\text{if~} c : \phi \text{~and~} d : \psi \\
     inl(c)  &: &  (\phi \Lor \psi)            ~~~~~ \text{if~}  c : \phi \\
     inr(d)  & : & (\phi \Lor \psi)            ~~~~~ \text{if~}  d : \psi \\
     c       & : &    (\phi \Imp   \psi)        ~~~~ \text{if~whenever~ }  d : \phi \text{~then~} c(d) : \psi \\
     (\lambda x: S) c & : & \Forall{x: S}{\phi}   ~~~\text{if~} \text{forall~} s \in \Dom{S}, c[s/x] : \phi[s/x] \\
     (s, c)    & : &  \Exists{x: S}{\phi}   ~~~  \text{if~} s \in \Dom{S} \text{~and~} c : \phi[s/x]
\end{eqnarray*}
In particular, logical conjunctions of formulas are realized by pairing, and logical disjunction is realized by elements of disjoint sum, depending on whether the left- or the right-hand side of the disjunction has been realized. Logical implication is realized by a function which produces evidence of the conclusion for each possible assumption. Given the definition $\phi \Imp \False$ for the logical negation of $\phi$, the realizability rule for logical negation is a special case of the one for implication.
Finally, universal quantification is realized by means of a function which produces evidence for all possible instantiations, and existential quantification is realized by a pair consisting of a witness $s$ together with an evidence that this witness indeed satisfies the formula in the body of the existential quantification. 

One way of implementing the BHK explanation of intuitionistic truth for arithmetic is to associate with each formula $\phi$ some collection of numerical codes for algorithms which could establish the constructive truth of $\phi$\@. Kleene's realizability semantics of intuitionistic arithmetic, in particular, assigns such a number $e$ that realizes  $\phi$ by induction on the syntactical structure of the sentence~\cite{kleene1973realizability}\@. 
An arbitrary formula is realizable if some number realizes its universal closure. In this way, the number $e$ realizes $\phi \Land \psi$ if $e$  codes a pair $(c,d)$ such that $c$ realizes $\phi$ and $d$ realizes $\psi$\@. 
Likewise, $c$ realizes $\phi \Imp \psi$ if, whenever $d$ realizes $\phi$,
then the $c$-th partial recursive function is defined at $d$ and its value realizes $\psi$\@. 
Thus, Kleene's realizability semantics encodes a proof checker for deciding if $c$ realizes a given formula $\phi$\@. 

There is a Gentzen-style proof system for Cyberlogic with sequents of the form $\Gamma \vdash c: \psi$, where $\Gamma$ is a finite set of assumptions of the form $x_i : \phi_i$\@. 
This proof system admits cut elimination,  
and attestation are proven to distribute with logical connectives as follows~\cite{nigam2021proof}\@.
     \begin{eqnarray*}
    \Attest{K}{\phi \Land \psi} & \Bimp & \Attest{K}{\phi} \Land \Attest{K}{\psi} \\
    \Attest{K}{\phi} \Lor \Attest{K}{\psi} & \Imp &  \Attest{K}{(\phi \Lor \psi)} \\
    (\Attest{K}{\phi \Imp \psi}) & \Imp & (\Attest{K}{\phi}) \Imp (\Attest{K}{\psi})\\
    \Attest{K}{(\nabla x:S)\, \phi)} & \Bimp & (\nabla x:S)\,\Attest{K}{\phi} 
                                     \text{~~~for~} \nabla \in \{\exists, \forall\}
    \end{eqnarray*}
Cyberlogic as investigated in~\cite{nigam2021proof} includes an additional knowledge modality $\Knows{\mathcal{Q}}{\phi}$,
for a set $\mathcal{Q}$ of principals,
with the intended meaning that $\phi$ is derivable from the collection of local knowledge bases of the principals $\mathcal{Q}$\@. 
Therefore, the operator $\Knows{\mathcal{Q}}{\phi}$ may be read as “principals in $\mathcal{Q}$ jointly (may) know $\phi$”\@.
In case $\mathcal{Q}$ is the emptyset, we refer to $\phi$ as {\em common knowledge}\@.

The following formula, for example, specifies that for $K$ to attest $\phi$, both $K_1$ and $K_2$ need to 
attest to $\phi$ using only their local knowledge:
   $$
       \Knows{\{K_1\}}{\Attest{K_1}{\phi}} \Land
       \Knows{\{K_2\}}{\Attest{K_2}{\phi}}
           \Imp \Attest{K}{\phi}
   $$
Another use of the knowledge modality is to specify that some evidence should only be trusted if derived 
from trusted sources. 
Consider, for example, three principals $\mathcal{P}=\{K_T,K_U,K\}$ where $K$ trusts evidence from $K_T$, but 
not all evidence from $K_U$\@. 
Then the following clause
   $$
      \Knows{\{K, K_T\}}{\Attest{K}{critical(ok)}}
      \Land
       \Knows{\mathcal{P}}{\Attest{K}{nonCritical(ok)}}
       \Imp
         \Attest{K}{\Knows{\emptyset}{all(ok)}}
   $$
specifies that $K$ can attest that everything is $ok$ as a common knowledge, i.e. $\Knows{\emptyset}{all(ok)}$,  
if all the non-critical and critical elements are $ok$\@. 
However, the check of critical parts can only be performed by principals $K$ trusts, namely $K$ itself or $K_T$\@.
In particular, information from $K_U$’s knowledge may not be used in the proof of $critical(ok)$\@.

The knowledge modality in Cyberlogic extends the one in  linear authorization logics~\cite{garg2006linear,nigam2014framework} in 
that it  supports the notion of knowledge shared by multiple principals. 
There are a number of interesting properties on knowledge expressible in Cyberlogic (see~\cite{nigam2021proof})\@:\footnote{
Notice that, strictly speaking,  the logic in \cite{nigam2021proof} does not include quantification over principals.
}
     \begin{eqnarray*}
     \Knows{\mathcal{Q}}{\phi} & \Imp  &\phi \\
     \Knows{\{K\}}{\phi} & \Imp &  \Attest{K}{\phi} \\ 
     \Knows{\mathcal{Q}_1}{\Knows{\mathcal{Q}_2}{\phi}} & \Imp &   \Knows{\mathcal{Q}_1 \cup \mathcal{Q}_2}{\phi}
     \end{eqnarray*}
On the other hand, not every fact is known to a set of principals,
attestation of a formula does not imply that the formula is known by the principal, 
and the knowledge modality does not commute.
     \begin{eqnarray*}
                   \phi     & \not\Imp  & \Knows{\mathcal{Q}}{\phi} \\
       \Attest{K}{\phi}     & \not\Imp  & \Knows{\{K\}}{\phi} \\
       \Knows{\mathcal{Q}_1}{\Knows{\mathcal{Q}_2}{\phi}} & \not\Imp  &  \Knows{\mathcal{Q}_2}{\Knows{\mathcal{Q}_1}{\phi}}
     \end{eqnarray*}
Knowing some attestation is different from attesting to this knowledge (for a detailed discussion see~\cite{nigam2021proof}).
    \begin{eqnarray*}
    \Knows{\mathcal{Q}}{\Attest{K}{p(t_1,\ldots, t_n)}} 
        & \not\Imp  & 
        \Attest{K}{\Knows{\mathcal{Q}}{p(t_1,\ldots, t_n)}}
    \end{eqnarray*}
The examples above demonstrate the expressive power of the knowledge modality in Cyberlogic,
and a Cyberlogic interpreter with attestation and knowledge modalities has been described previously~\cite{nigam2021proof}\@. 
In order to keep the presentation as succinct as possible, however, we will not explicitly mention
the case of Cyberlogic policies with knowledge modalities in the remainder of this exposition.
     

\chapter{Distributed Proof Search}     \label{sec:proof.search}

Given a logical context $\Gamma$ of the form $x_0: \phi_0, \ldots, x_n: \phi_n$, for some $n \in Nat$, and a formula $\psi$, the problem of {\em proof search} is to determine a proof $c$ such that
$\Gamma \,\vdash\, c : \psi$\@. 
Clearly, proof search for first-order intuitionistic logic, and therefore also for Cyberlogic, is undecidable. 
The problem of checking if some given proof actually proves some statemeent, on the other hand, usually is of 
low polynomial complexity.

For distributed proof search, we assume a certain number of principals $K$ in Cyberlogic\@.  
We also assume that each of these principals has a local logical context $\Gamma(K)$\@.
These local contexts form the {\em policies} of 
the principals $K$, as they form the basis for $K$'s own (local) deductive proof search capability. 
In this way, proof search may proceed locally until there is a subgoal which needs to be handed-off to another principal. 
We refer to this integrated process of local deduction and the communication of subgoals and (partial) solutions between 
authorities as {\em distributed proof search}\@. 

\begin{figure}[t]
 
 Policy $\Gamma(A)$ 
 \begin{eqnarray*}
        a_1 & : & \Attest{A}{isHospital(C)}  \\
        a_2 & : & (\forall X: Physician, Y: Patient)\, 
                       isPhysicianOf(X,Y) \\
            &   & ~~~ \Imp \Attest{A}{readMedRec(X,Y)} \\
       a_3  & : & (\forall X: Physician, Y: Patient, Z: Hospital) \\
            &   & ~~~\Attest{A}{isHospital(Z)} \Land      
                     (\Attest{Z}{isPhysicianOf(X,Y)}) \\
            &  &  ~~~~~~     \Imp   \Attest{A}{isPhysicianOf(X,Y)}\\
       a_4  & : & (\forall H, Z_1, Z_2: Hospital)\, Z_1 \neq Z_2 \Land Z_1 \neq A \Land Z_2 \neq A \\
            &   & ~~~\Land \Attest{A}{isHospital(Z_1) \Land isHospital(Z_2)}  \\
            &   & ~~~\Land (\Attest{Z_1}{isHospital(H)} \Lor           \Attest{Z_2}{isHospital(H)})  \\  
            &   & ~~~~~~\Imp \Attest{A}{isHospital(H)}
\end{eqnarray*}
Policy $\Gamma(B)$ 
 \begin{eqnarray*}
       b_1 & : &  \Attest{B}{isHospital(A)} \\
       b_2 & : &  \Attest{B}{isHospital(B)} \\
       b_3 & : & \Attest{B}{isPhysicianOf(Alice, Peter)} 
 \end{eqnarray*}
 Policy $\Gamma(C)$ 
 \begin{eqnarray*}
       c_1 & : &  \Attest{C}{isHospital(B)}
 \end{eqnarray*}
 
    \caption{Scenario of hospital policies for accessing medical records.}
    \label{fig:hospital}
\end{figure}
    
We illustrate distributed proof search by means of a simple example for accessing medical records. There are three hospitals with names $A$, $B$, $C$ with corresponding authorities\@. 
The access policies $\Gamma(A)$, $\Gamma(B)$, and $\Gamma(C)$ for the three hospitals are depicted in Figure~\ref{fig:hospital}\@. 
For instance, $a$ attests to the facts that $b$ and $c$ are hospitals, and $a_2$ encodes the rule that every physician of some patient can read her medical records. 
Assumption $a_3$ allows $A$ to {\em delegate} the decision on the physician-patient relationship to some hospital $Z$; more precisely, if hospital $Z$ attests to a certain physician-patient relationship then $A$ attests to the truth of this relationship.

Now, assume that some evidence needs to be constructed that indeed physician $Alice$ can access the medical records of patient $Peter$\@.
    \begin{eqnarray*}
     ?_0 & : & \Attest{A}{readMedRec(Alice,Peter)}
    \end{eqnarray*}
Here, $?_0$ is a (meta-)variable for the evidence which needs to be constructed. 
We use the assumptions of the individual policies in Figure~\ref{fig:hospital} to iteratively collect constraints on these meta-variables. 
In a first step, we apply hypothesis $a_2$ to $Alice$ and $Peter$ to obtain the new subquery.
    \begin{eqnarray*}
       ?_1  & : &   \Attest{A}{isPhysician(Alice, Peter)}
    \end{eqnarray*}
Since $?_0 \equiv a_2(Alice)(Peter)(?_1)$ one can construct an evidence of the top query from a solution for $?_1$\@. 
In proof search these kinds of reductions are called {\em backward chaining}, and their application relies on matching (possibly modulo theories) the current query with the conclusion of a universally quantified implication. 
In this way, the correspondingly instantiated premises of this implication serve as the new subqueries.
In a next step, we backchain on assumption $a_3$ to obtain
   \begin{eqnarray*}
       ?_2  & : &   \Attest{A}{isHospital(Z)}\\
       ?_3  & : &   \Attest{Z}{isPhysicianOf(Alice, Peter)}\\
    \end{eqnarray*}
with $?_1 \equiv a_3(Alice)(Peter)(Z)(?_2)(?_3)$\@. 
Thus, one needs to find a hospital $Z$ which attests to this physicial-patient relationship. 
This is a fact in the policy of hospital $B$, and consequently $?_3 \equiv B$ and $Z \equiv B$\@. 
Thus, we are reduced to show that $A$ can attest that $B$ indeed is a hospital. 
Backchaining on rule $a_4$ we need to find two hospitals distinct of $A$ which are attesting that $B$ indeed is a hospital.
  \begin{eqnarray*}
                  ?_5 & : & Z_1 \neq Z_ 2 \Land Z_1 \neq A \Land Z_2 \neq A \\
                  ?_6 & : &   \Attest{A}{isHospital(Z_1)} \Lor   
                              \Attest{A}{isHospital(Z_2)}\\
                  ?_7 & : &   \Attest{Z_1}{isHospital(B)} \Land  
                                  \Attest{Z_2}{isHospital(B)}\\
    \end{eqnarray*}
Here, $?_2 \equiv a_4(B)(Z_1)(Z_2)(?_5)(?_6)(?_7)$\@.
With $Z_1 \equiv B$ and $Z_2 \equiv C$ we obtain $?_6 \equiv inr(a_1)$,
$(?_7 \equiv (b_2, c_1)$, and $?_5$ is obtained from the corresponding background theory ("all hospitals are not equal")\@. 
This concludes the construction, and evidence $?_0$ for the original query is obtained by back-substitution: 
    \begin{eqnarray*}
       &&   a_2(Alice)(Peter) \\
       &&   ~~~(a_3(Alice)(Peter)(B) \\
       &&   ~~~~~~(a_4(B)(B)(C) \\
       &&   ~~~~~~~~~(\_)(inr(a_1)(b_2,c_1)) \\
       &&   ~~~~~~(b_3))
    \end{eqnarray*} 
We left a "hole" $(\_)$ for the proof term for the query $?_5$ as obtained from the background theory\@.
Decision procedures for certain background theories may be used to fill in these kinds of "holes" when needed.
This may result in drastically reduced sizes of proof terms, but at the expense of increased effort for checking evidence.

\begin{figure}[t]
Policy $\Gamma(CA)$ 
 \begin{eqnarray*}
        ca_1 & : & (\forall x: Object)\, \Delegate{CA}{HMO}{isHospital(x)}
\end{eqnarray*}
Policy $\Gamma(HMO)$ 
 \begin{eqnarray*}
       hmo_1 & : & (\forall K: Principal,\, x: Object)\, (K = A \lor (K = B \lor K = C)) \Imp \\       
             & & ~~~\Attest{K}{isHospital(x)} \Imp \Attest{HMO}{isHospital(x)}
 \end{eqnarray*}
    \caption{Example for hospital policies and delegation.}
    \label{fig:hospital.delegate}
\end{figure}

Incremental synthesis of proof terms in intuitionistic logic by means of back-chaining is supported by 
proof development systems such as Coq~\cite{paulin2011introduction}\@.
In our case, however, the proof search is distributed between several collaborating authorities. 
Consider, for example, the situation above where $A$ needs to find a hospital $Z_1 (\neq A)$ with $\Attest{Z_1}{isHospital(B)}$\@. 
Clearly, given its policy $\Gamma(A)$, the authority $A$ can not progress on solving this subquery, and 
consequently initiates a broadcast to applicable authorities. 
In the example, above both $B$ and $C$ respond successfully, and $A$ is able to attest that both respondents are indeed hospitals. 
Alternatively, $a$ may selectively only query authorities for which it is able to attest that they indeed are hospitals.
Similar strategies for distributed proof search, in particular, the interaction between local and global proof search, have recently been studied, for example, in~\cite{nigam2021proof,nigamproof}\@.
    
The proof term as generated above may be used as evidence that $Alice$ has access to the medical records of $Peter$\@.
Indeed, a gatekeeper, physical or not, of the requested medical records may verify access rights by checking that the presented proof term supports the requested access. 
These kinds of proof checks can be performed in low polynomial time in the size of the proof term. 

There are however a number of disadvantages to this style of proof-carrying authorization in real-world applications. 
First, proof checking requires access to all policies.  
Despite all the wishes for transparency on policies, this does not seem to be a realistic proposition. 
For example, it seems highly unlikely that all authorities involved in a visa application process make all of their decision policies widely accessible. 
Second, proof terms represent tree-like derivations which may be growing arbitrarily large. 
There are some techniques of shrinking proof trees. For example, proof terms for decidable fragments (for example, background theories such as Presburger arithmetic) may be omitted, since these proofs can always be reconstructed. 
Other proof-theoretic techniques for compact representations of proofs such as {\em midsequents}~\cite{gallier2015logic} and {\em focusing}~\cite{nigam2021proof}) are also  based on trading space for representing proofs with the time for proof checking~\cite{miller2015foundational}\@. 
We will address these issues in some more detail in~\ref{sec:crypto.proofs}\@.

\chapter{Expressiveness}  \label{sec:expressiveness}

\section{Authorization and Delegation}  \label{sec:delegation}

The attestation $\Attest{K}{\phi}$  does not logically imply the validity of $\phi$\@. 
Such a validity only follows when principal $K$ possesses {\em authority} over $\phi$\@; 
namely:
$$
     (\Attest{K}{\phi}) \Imp \phi\mbox{\@.} 
$$
Consider, for example, a principal $\mathrm{A}$ for deciding the theory of linear arithmetic. {\rm Trust} on principal $\mathrm{A}$ to be correct in the sense that it only attests valid arithmetic formulas may be expressed through the authority $\Attest{A}{\phi} \Imp \phi$ for all $\phi$ in the language of linear arithmetic. 

Such an authority of a key must itself be attested by a certification authority.
Let's assume a single 'top-level' certification authority $CA$ with the assumption that anything attested by this certification authority is valid. Thus, for all $\phi$\@:
$
(\Attest{CA}{\phi}) \Imp \phi 
$\@.
A principal $K$ can selectively authorize another principal $L$ through
$$
    \Attest{K}{(\Attest{L}{p(x_1,\ldots, x_n))}\Imp p(x_1,\ldots, x_n)}\mbox{\@.}
$$
This kind of  delegation of the authority over certain formulas $p(x_1,\ldots, x_n)$  from a principal $K$ to a principal $L$ may also be denoted by
$\Delegate{K}{L}{\phi}$\@.
In this way, the certification authority $CA$ in Figure~\ref{fig:hospital.delegate} delegates the authority on $isHospital$ to a principal $HMO$, which itself chooses to attest to $isHospital$ only if one of its hospital principals $A$, $B$, or $C$ does so.

Likewise, the authority of $visa(\_)$ in our running example on issuing visa certificates may be selectively delegated to certain consulates by the State Department. 
Such a delegation might also be realized through a chain of delegation.
And, indeed, indirect delegation is definable in Cyberlogic as follows:
   $$
    \Attest{K}{(\forall M: Principal) \Attest{M}{\phi} \Imp (\Attest{L}{(\Attest{M}{\phi}) \Imp \phi}) \Imp \phi}\mbox{\@.}
   $$
Moreover, indirect delegation can be iterated to an arbitrary finite depth.
Initial versions of Cyberlogic distinguished between direct and indirect attestations, where direct attestations were used to restrict the capability of further delegation.
However, the proof theory of Cyberlogic with direct and indirect attestations does not admit cut-elimination~\cite{reis2019observations}, and has therefore been omitted\@. 

\section{Timed Attestation} \label{sec:time}

Time is often needed to establish that a piece of evidence was produced before or after a certain point in time. 
The $Time$ domain for representing instances of time is infinite, it admits an equality relation $=$, and it is 
totally ordered by $<$; otherwise, the $Time$ domain is uninterpreted. 
For the certification of time we assume a {\em trusted time source} $T$ of sort $Principal$ and a unary predicate $time(t)$ for attesting that a specific time instant $t$ has passed; that is:
\begin{eqnarray*}
     c & : & \Attest{T}{time(t)}\mbox{\@,}
\end{eqnarray*}
where $c$ certifies that time $t$ has elapsed if $\Decr{\Pub{T}}{c}$ yields $time(t)$\@.

The following definitions are used for specifying whether some point in time $t$ has elapsed or not, or if it is current. 
\begin{eqnarray*}
      past(t)  & :=  & (\exists s: Time)\, s > t \Land\ \Attest{T}{time(s)} \\
      future(t)   & :=  & (\forall s: Time)\, s \geq t \Imp \Lneg(\Attest{T}{time(s)}) \\
      curr(t)     & := &  \Attest{T}{time(t)} \land (\forall s: Time)\, \Attest{T}{time(s)}                              \Imp  s \leq t
\end{eqnarray*}
Time-stamped certificates are embedded in other certificates to establish that a piece of evidence was produced following time $t$\@. \begin{eqnarray*}
     \AttestAfter{K}{t}{\phi} & := & \Attest{K}{\phi} \Land \Attest{T}{time(t)}
\end{eqnarray*}
Therefore, $c$ realizes this timed attestation if $\Decr{\Pub{K}}{c}$ yields a representation of a pair $(c, d)$ such that $c$ certifies $\Attest{K}{\phi}$ and $d$  certifies $\Attest{T}{time(t)}$\@.
The situation here is similar to hijackings scenes in the movies, where the culprit sends a picture of a hostage together with a recent issue of a newspaper to provide evidence of her relative well-being at a point of time after the newspaper has been issued.

To demonstrate that a claim $\Attest{K}{\phi}$ was produced prior to time $t$, we once again rely on the trusted time source $T$ to timestamp the claim.
The attestation $\AttestBefore{T}{t}{\phi}$ indicates that the trusted time source $T$ attests to $\phi$ being verifiable prior to time $t$\@.
In this case, the certificate $c$ is evidence for $\AttestBefore{T}{t}{\phi}$ if $\Decr{\Pub{T}}{c}$ yields $(\phi, t)$\@.
Since $T$ is a trusted time source, its timed attestations can be taken as valid, so that
$
 \AttestBefore{T}{t}{\Attest{K}{\phi}} \Imp \AttestBefore{K}{t}{\phi} 
$\@.
%

Finally, temporal modalities are definable in Cyberlogic using timed attestations ($t$ is of sort $Time$)\@.
\begin{eqnarray*}
    \AttestEventually{K}{\phi}(t) & := & \AttestAfter{K}{t}{\phi} 
\end{eqnarray*}
Thus, 
$\AttestEventually{K}{\phi}$ holds at time $t$ if $K$ attests to $\phi$ at some latter point of time.
These kinds of encodings are used for defining a temporal logic by pointwise lifting the logical connectives to the time domain, and for deriving corresponding inference rules in the usual way. 
Past-time temporal operators are encoded in a similar manner.
Moreover, interval-based temporal operators such as  $ \AttestGlobalInterval{t_1}{t_2}{K}{\phi}$, which
expresses that $k$ always attests to $\phi$ in the time interval $[t_1, t_2]$ and, more generally, Allen-style~\cite{allen1989moments} operators such as $\mathrm{Before}$, $\mathrm{After}$, $\mathrm{During}$, and $\mathrm{Overlaps}$ are also definable\@.

\section{Revocation of Authority}\label{sec:revocation}

Timestamping can be used to ensure that revocable certificates have not been revoked at a given time.
This can be done by ensuring that the authority granted by the certification authority $CA$ is fresh, as given by
$$
      \AttestBefore{T}{t}{(\Attest{CA}{(\Attest{K}{\phi}) \Imp \phi)}}\mbox{\@.}
$$
However, this mechanism does not specify that the authority is revocable. 
A revocable delegation of authority from $K$ to $L$ can be specified as
$$
           \Attest{K}{(\Attest{L}{\phi(s) \Land \Attest{K}{notRevoked(L, t) \Land s < t}} \Imp \phi(s))}\mbox{\@.}
$$
This formula asserts that if $K$ attests that $L$'s authority has not been revoked up to time $t$, then any attestation by $L$ of the statement $\phi(s)$ for $s < t$ is tantamount to an attestation by $K$ of $\phi(s)$\@.

This simple mechanism demonstrates how revocation of authorities can be encoded in core \Cyberlogic\@.
Clearly, more refined revocation mechanisms are possible, for example by restricting the authority over certain formulas or to allow for undoing revocations. 

\section{Nonces} \label{sec:nonces}

Another mechanism for ensuring the freshness of evidence is through the use of {\em nonces}, which can be thought of as unique numbers which are hard or impossible to guess.
We therefore assume a trusted principal $N$ for generating nonces $n$  by means of asserting:
   $$ \Attest{N}{nonce(n)}\mbox{\@.} $$
Uniqueness of the nonces is expressed by the assumption
   $$ (\forall n, m: Nonce, s, t: Time)\, 
         s \neq t \land
         \AttestAt{N}{s}{nonce(n) } \land
          \AttestAt{N}{t}{nonce(m) } \Imp n \neq m
   \mbox{\@.} 
   $$
where $\AttestAt{N}{t}{nonce(.)}$ denotes attestation  of $nonce(.)$ at time point $t$\@. 

A common mechanism in protocols is that one participant in a protocol  challenges another participant to produce evidence with a newly generated number, the nonce, so as to ensure recency of the evidence. 
These kinds of mechanisms have previously been formalized in terms of logic in the context of specifying security protocols~\cite{durgin2004multiset}\@. 
In this way, we will also be using logical instantiation rules for generating nonces while executing security protocols (see Section~\ref{sec:needham.schroeder})\@.

\section{Distibuted Logic Programs} \label{sec:execution}

Transactions and protocols are implemented as distributed logic programs. 
The policies in Figure~\ref{fig:hospital}, for example, can easily be expressed as universally closed
Horn formulas of the form $\phi_1 \Land \ldots \Land \phi_n \,\Imp\, \phi$, where each $\phi_i$ and $\phi$ is either an atom or an {\em atomic attestation} of the form $\Attest{k}{\psi}$, where $\psi$ is an atom\@. 
A corresponding query is an atom or atomic attestation such as $\Attest{a}{readMedRec(Alice, Peter)}$ (see Section~\ref{sec:proof.search}),
and the logical derivation of this query from the union of individual policies is based on an operational reading of Horn clauses, called  {\em backchaining}, for
successively reducing  queries $\phi$ to subqueries $\phi_1, \ldots, \phi_n$\@. 

The example in Section~\ref{sec:proof.search}) also demonstrates the three basic steps of the distributed execution of Cyberlogic programs 
for proof search of queries, as, in each execution step, 
either (1) a query is decomposed into subqueries, 
or (2) a query is sent from a supplicant to an appropriate authority, 
or (3) there is a reply to a query in the form of a (partial) proof sent back to the supplicant. 
The steps (2) and (3) for communicating queries and partial solutions between principals may also be 
decoupled further using a distributed coordination mechanism such as Linda~\cite{gelernter1992coordination}\@. 
These kinds of blackboard architectures also support queries such as $(\exists X: Principal)\,\Attest{X}{p(42)}$ when an adequate addressee for solving this query has not been 
determined by the principal issuing this query.
These generalized queries are reminiscent of multicast broadcasting.

Cyberlogic~programs are based on a slight extension of Horn formulas, which are called (first-order) {\em hereditary Harrop formulas}\@.
These formulas extend the logic of Horn clauses by permitting implication and universal quantification in queries, and they
consist of {\D}-formulas (programs) and {\G}-formulas (queries) \cite{miller1991uniform}\@.
     \begin{eqnarray*}
     \D   & ::=& \Atom \; | \; \G \Imp \Atom \; | \;  (\forall x) D  \; |  \; D \Land D \\
     \G   & ::=& \Atom 
                  \; |  \; \G \Land \G 
                  \; |  \; \G \Lor \G 
                  \; |  \; (\exists x)\,\G  
                  \; |  \;  \D \Imp \G 
                  \; |  \; (\forall x) \G 
     \end{eqnarray*}
Here, {\Atom} is any atom or an atomic attestation modality.
These formulas define a logic programming language in that
a {\G}-formula can be thought of as a query (or goal), a finite set of {\D}-sentences constitutes a program, 
and the process of answering a query consists of constructing an intuitionistic proof of the existential closure of the query from the given program. 
In this way, hereditary Harrop formulas constitute an abstract logic programming language~\cite{miller1991uniform}, since the declarative reading of the logical connectives coincides with the search-related reading\@.
A query of the form $\D \Imp \G$, in particular, can be interpreted as an instruction
to augment the program with $\D$ in the course of solving $\G$,
and a query of the form $(\forall x)\,\G$ can be interpreted as an instruction to generate a new name and to use it for $x$ in the course of solving $\G$\@.

This operational reading of Cyberlogic formulas yields a proof procedure for the logic of hereditary Harrop formulas~\cite{nadathur1993proof}\@.
This proof procedure is based on searching for uniform proofs~\cite{miller1991uniform} in the logic of hereditary Harrop formulas, and uses concepts such as dynamic Skolemization \cite{fitting2012first} and unification of terms embedded under arbitrary sequences of quantifiers for controlling the search for instantiations of existentially quantified goal formulas.
This proof procedure can be extended to higher-order hereditary Harrop formulas, which are used in the higher-order logic programming language $\lambda$Prolog.
Indeed, early Cyberlogic interpreters have been realized as deep embeddings in $\lambda$Prolog~\cite{bernat2006first}\@.  

A large number of protocols and policies may already be encoded as variants of Horn-clause logic programs.
In this way, we have identified a Horn-clause like fragment of Cyberlogic, and proposed a (focused) proof systems for sound and complete proof search~\cite{nigam2021proof}\@.  

Datalog is even less expressive as it restricts Horn-clause logic programs to atoms of the form $p(a_1,\ldots,a_n)$, where each $a_i$ either is 
 a variable or a constant symbol. 
Indeed, Datalog is attractive in that it supports the evaluation of recursive programs, it may be extended with built-in predicates (which evaluate variable-free atoms)~\cite{li2003datalog}, supports (stratified or well-founded) negation~\cite{zhou1995fixpoint}, and may be  executed in a distributed manner~\cite{abiteboul2010distributed}\@. 
Datalog with stratified negation, in particular, has been instrumental in expressing a number of non-trivial policies~\cite{lam2009formalization}\@. 
ETB Datalog, in addition, has well-defined denotational\footnote{ETB Datalog uses an unbounded Herbrand universe in contrast to vanilla Datalog} and distributed operational semantics~\cite{cruanes2014semantics}, it generalizes built-in predicates to interpreted predicates, and ETB Datalog may generates certificates. 
Finally, the Cyberlogic interpreter in~\cite{nigam2021proof} is unique in that it supports (focused) execution of programs with both attestation and knowledge modalities.

\chapter{Cryptographic Protocols} \label{sec:needham.schroeder}

We are using the infamous Needham-Schroeder~\cite{lowe1996breaking} authentication protocol for
demonstrating the expressive power of Cyberlogic and distributed proof search for formalizing and executing cryptographic protocols.
The textbook presentation of the Needham-Schroeder protocol specifies the sequence of messages between senders and intended recipients.
     \begin{eqnarray*}
     A \to B & : &  \{A, N_a\}_{\Pub{B}}\\
     B \to A & : &  \{B, N_a, N_b\}_{\Pub{A}}\\
     A \to B & : &  \{N_b\}_{\Pub{B}}
     \end{eqnarray*}
In a first step, principal $A$ sends $B$ an encrypted message containing both the name of the sender together with a nonce $N_a$, which has been generated by $A$ for this purpose\@. 
Principal $B$ responds by generating a nonce $N_b$, conjoining it with the nonce $N_a$ as obtained by $A$ together with its name $B$, and sending the encrypted message to $A$\@. 
In a third step, $A$ acknowledges receipt of this message to $B$\@.
The goal of this protocol is to generate a secret number, which is known to the principals $a$ and $b$, but nobody else.
From this informal description it should become evident that the textbook description above on the Needham-Schroeder sequence of message exchanges does not make the details for executing the protocol explicit. A complete formalization in logic of these kinds of security protocols  has been developed, for example, in~\cite{millen2000protocol}\@. 

It turns out that the hereditary Harrop formulas of Cyberlogic are expressive enough to succinctly formalize cryptographic protocols such as Needham-Schroeder authentication. 
In fact, principal $A$ initiates a Needham-Schroeder protocol round by querying principal $B$ based on the Cyberlogic formula below, and $B$ responds using the same Cyberlogic program $NSP(A, B)$, which is defined as
    \begin{eqnarray*}
    &&(\forall x_a: Nonce)\\
    &&  ~~~ ((\forall x_b: Nonce)\, \Attest{B}{msg_3(x_b)} \,\Imp\, \Attest{A}{msg_2(B, x_a, x_b)}) \\
    &&  ~~~~~~ \Imp\, \Attest{B}{msg_1(A, x_a)}\mbox{\@.}
    \end{eqnarray*}
Thus, $NSP(A, B)$ represents one round of the Needham-Schroeder protocol between principals $A$ and $B$\@. 
In initiatiating such a round, $A$ first chooses a (constant) nonce $N_a$\footnote{To keep this exposition as simple as possible $A$ just chooses an arbitrary integer without any considerations on the freshness of this choice.} and queries $B$ to provide evidence $?_0$ that it attests to $msg_1(A, N_a)$\@.
    \begin{eqnarray*}
     h_a   & : &    ((\forall x_b: Nonce)\, \Attest{B}{msg_3(x_b)} \,\Imp\, \Attest{A}{msg_2(B, N_a, x_b)}) \\
     ?_0   & : & \Attest{B}{msg_1(A, N_a)}
    \end{eqnarray*}
Now, $B$ backchains based on its program and selects, as above, a nonce $N_b$ to obtain the subquery in its extended local context. 
    \begin{eqnarray*}
       h_b  & : &  \Attest{B}{msg_3(N_b)} \\
       ?_1  &   &  \Attest{A}{msg_2(B, N_a, N_b)} 
    \end{eqnarray*}
Principal $B$ now queries $A$ to provide such an evidence $?_1$ for its attestation to $msg_2(B, N_a, N_b)$\@.
Principal $A$ responds to this query by backchaining on program $h_a$ after instantiating it with the nonce $N_b$ as obtained through the message $msg_2$ from $B$\@.
This results in the subquery 
     \begin{eqnarray*}
     h_a   & : &    ((\forall x_b: Nonce)\, \Attest{B}{msg_3(x_b)} \,\Imp\, \Attest{A}{msg_2(B, N_a, x_b)}) \\
     ?_2   & : &    \Attest{B}{msg_3(N_b)}
    \end{eqnarray*}
which is communicated and delegated to $B$\@. This query is solved trivially by hypothesis $h_b$\@.

This concludes the distributed execution for solving $A$'s query. 
Notice that the subqueries as communicated back-and-forth between the principals $A$ and $B$ in the derivation above are exactly the plain text versions of the messages of one round of the Needham-Schroeder protocol, and they occur in exactly the same order. 
If communication channels between principals is insecure, we may, following the Needham-Schroeder protocol above, only communicate queries encrypted by the public key of the intended recipient. 

The construction of the proof term for the original query from its subqueries is immediate from the realizability semantics of Cyberlogic.
Such a proof term is useful for verifying that the result has been obtained in a manner compliant with the Needham-Schroeder protocol.
It does not, however, directly entail the crucial secrecy property of the Needham-Schroeder protocol itself.
Cyberlogic's knowledge modality may be used for expressing  secrecy properties of the Needham-Schroeder protocol.
   \begin{eqnarray*}
   NSP(A,B) & \Imp & (\exists N: Nonce)\,\, \Knows{\{A\}}{nonce(N)} \Land \Knows{\{B\}}{nonce(N)} \\
        &   &   ~~~~~~ \Land (\forall K: Principal)\, (K \neq A \Land K \neq B) \\
        &   &   ~~~~~~~~~~\Imp \Lneg{\Knows{\{K\}}{nonce(N)}} 
   \end{eqnarray*}
That is, after one round of the Needham-Schroeder protocol between principals $A$ and $B$, these principals share a secret which nobody else knows.
These kinds of secrecy properties for cryptographic protocols are established by means of inductive secrecy invariants~\cite{millen2000protocol}\@.

The above derivation demonstrates that the operational semantics of the respective contexts of the principals determines the construction of the next subquery / message at any specific point of time. 
The semantics of the logical connectives therefore provide all the necessary details and the right kind of scoping of variables for continuing executing the Needham-Schroeder protocol.
In fact, these local contexts can be thought of as continuations. 
For sequential protocols such as Needham-Schroeder there is no {\em don't know} non-determinism in the operational proof search as the choice of the nonces is arbitrary. 
And, therefore, distributed proof search might simply be realized by interacting coroutines for each principal which 
are switching control according to continuations based on local proof search.

\chapter{Privacy of Policies}\label{sec:crypto.proofs}

Evidence checking as outlined above requires a complete proof tree and also access to all policies.
These are strong assumptions, indeed, as tree-like proof terms may grow arbitrary large, 
and not all principals are ready to be completely transparent with respect to their policies.
The size of the proof trees may simply be reduced by node sharing, for example, by maintaining
the hashs of proof nodes in a tamper-proof lookup table. 
For supporting evidential transactions betweeen principals with private policies, 
we propose the use of {\em functional encryption} schemes~\cite{boneh2011functional}\@. 

A proof/evidence checker for Cyberlogic is a function $check_\Gamma(e, \phi)$ which returns $ok$ if $\Gamma \vdash e : \phi$, and $nok$ otherwise.
A proof checker for the principal $HMO$ with the logical context $\Gamma(HMO)$ in Figure~\ref{fig:hospital.delegate}, for example, is derived directly from the structure of the single policy rule $hmo_1$\@. 
    \begin{align*}
    & \mathrm{check}_{\Gamma(HMO)}(e, \phi) ~=\\
    &~~~\mathbf{cases}~\phi~\mathbf{of} \\
    &~~~~~~(\Attest{HMO}{isHospital(x)})~: \\
    &~~~~~~~~~\mathbf{cases}~\Decr{\Pub{HMO}, e}~\mathbf{of} \\
    &~~~~~~~~~~~~hmo_1(A)(x)(inl(\_))(d)~:~\mathrm{check}_{\Gamma(A)}(d, \Attest{A}{isHospital(x)}) \\
    &~~~~~~~~~~~~ \ldots \\
    &~~~~~\mathbf{otherwise}~:~nok
    \end{align*}
If evidence for attestations are hashed, then proof terms are obtained from a lookup table,
thereby avoiding the need to communicate large proof trees.
    
The definition of the proof checker above requires knowledge about all local policies involved in building up the evidence.
Therefore, if principals decide to keep respective policies secret then verifiable proof-carrying authorization as outlined above does 
not work anymore.
We demonstrate a possible way forward to keep such a checker secret and still be able to apply it for verifying specific authorities.
This approach is inspired by {\em functional encryption}~\cite{boneh2011functional}, which, given a function $f$ constructs a 
key pair $(\Pub{f}, \Prv{f})$ such that  $f(x)$ can be computed as $\Decr{\Prv{f}}{\Encr{\Pub{f}}{x}}$\@. 
In this way, the function $f$ may be evaluated by a third party (say, the "cloud") without 
revealing arguments $x$ to this untrusted party.

The situation here is slightly different as we are interested in keeping the function, namely the proof checker 
secret, whereas the argument $(c, \varphi)$, which includes a certificate $c$, may be public knowledge.
Therefore, given a checker $check_{\Gamma}$, which is only known to the principal with policy $\Gamma$, 
we assume the existence of key pairs $(\Pub{check_\Gamma}, \Prv{check_\Gamma})$
such that
    $$
      check_\Gamma(c, \phi) = \Decr{\Prv{check_\Gamma}}{\Encr{\Pub{check_\Gamma}}{(c,\phi)}}\mbox{\@.}
    $$
In this way, a certificate $c$ is verified by means of (a partial evaluation) of composing
encryption followed by decryption using the functional encryption pair for $check_{\Gamma}$, 
and without revealing the policy $\Gamma$\@. 
As such, our traveler presents a certificate to the border control which checks, 
using the functional key pair for the proof checker of the consulate $Cons42$ for 
permission of entry. 

The functional encryption scheme is also used for the {\em incremental maintenance} of policies and certificates. 
This requires adding a hash value $\Hash{\Gamma}$ to certificates and the maintenance of a tamper-proof log of
the publicly accessible table $\Hash{\Gamma} \mapsto check_\Gamma$, which is used for 
looking up the appropriate proof checker for the policy $\Gamma$\@. 
Whenever a principal updates its policy, a new entry to this lookup table is generated.
These changes may also require to recursively update all dependent policies.
Certificates which have been generated with outdated policies, can still be checked
as the corresponding functional key pair is being kept as part of the lookup table.

It is at least unusual to publish both the public and the secret key, 
and there is a chance that the checker and the actors' policies, which are supposed to be kept private,  might be reconstructed from this key pair. 
It should therefore be interesting to investigate the possibility of generating a single key for replacing proof checking with decryption.
Clearly, the checker might also be approximated inductively from observing its outputs, which needs to be restricted.

A possible way forward for ensuring that $\Gamma$ remains private is based on multi-input functional encryption~\cite{gordon2013multi,goldwasser2014multi}, which 
generalizes functional encryption to $n$-ary functions both in public-key and the symmetric-key settings. 
Here, a user in possession of a token $TK_f$ for an $n$-ary function $f$ and multiple ciphertexts $\Encr{.}{x_1}$, \ldots , $\Encr{.}{x_n}$ can 
compute $f(x_1,...,x_n)$ but nothing else about the arguments $x_i$\@.
We therefore consider the proof checker $check$ as the uncurried binary function $check(\Gamma, (c, \phi))$\@.
Now, a multi-input functional encryption scheme can be used for encrypting the two arguments $\Gamma$ and $(c, \phi)$ separately. 
The function $check$ can then be applied to these arguments without revealing anything about the arguments.
The point is that the encryption scheme ensures that the information in $\Gamma$ remains private. 
The function $check$ therefore may be published without revealing the policy $\Gamma$\@.


\chaptermark{Related Work} \label{sec:related.work}

The original design of Cyberlogic~\cite{ruess2003introducing} has been  influenced heavily  by the trust management view of  
authorization in open, large-scale, distributed systems~\cite{blaze1996policymaker, ellison1999spki, li2003delegation,chu1997referee}\@.
This view adopts a key-centric perspective of authorization, since public keys are treated as principals and direct authorization.
There is also a long tradition of encoding authorization policies in terms of logic programs (for example, ~\cite{li2003datalog,anderson2003constraint}), and proof-carrying authorization has also been proposed previously~\cite{appel1999proof}\@. 

The novelty of Cyberlogic is largely due to the combination of its three key ingredients: first,  the identification of public keys with authorities, second,  the specification of authorization policies as distributed logic programs, and third, the construction of verifiable and tamper-proof evidence by distributed proof search. 
This combination makes Cyberlogic expressive enough for encoding a number of all-important policy mechanisms such as revocation and delegation~\cite{li2003delegation}, which previously have been custom-built into specialized trust management systems or added in a rather {\em ad hoc} manner. 

Cyberlogic has found a number of different applications over the years.
For instance, Cyberlogic has been proposed for producing nonrepudiable {\em digital evidence} sound enough to withstand court challenges~\cite{gehani2009system},
as an enabling mechanism for accountable clouds~\cite{gehani2013accountable},
for generating safety cases~\cite{beyene2021carlan},
for specifying smart legal contracts~\cite{dargaye2018towards,dargaye2018pluralize} - 
since trusting the majority of system nodes in a blockchain does not need to be assumed any more
with verifiable and possibly proof-carrying computation,
for enabling accountability in federated machine learning~\cite{baracaldo2022towards, balta2021accountable} including verifiability, undeniable consent, auditability, and tamper-evidence, 
and, recently, also for assuring the trustworthiness of increasingly autonomous systems~\cite{ruess2022systems}\@.

DKAL is a declarative authorization language for distributed systems~\cite{gurevich2008dkal}, which shares the Cyberlogic vision of logic and proof-based authorization in distributed systems. 
The basic setup of Cyberlogic and DKAL are similar, in that DKAL also is  a  language  for  writing  the  policies  of  a  distributed  system’s 
interacting principals, each DKAL  principal  performs  its  own  computations  and  deductions, and the system has no central logic engine but merely provides an infrastructure supporting reliable communication between principals~\cite{blass2012introduction}\@. 
Also, each principal works with a knowledge base
that can be modified as a result of communications from other principals or as a result of deduction. 
Principals in DKAL, however are communicating by sending each other items  of  information,  called infons, whereas Cyberlogic principals communicate 
queries and answers to these queries in the process of distributed proof search. 
Also, the main aim of DKAL is to exceed the expressivity of the Datalog-based
SecPal~\cite{becker2010secpal} trust management system in a number of useful ways and yet to maintain feasible complexity bounds for answering authorization queries~\cite{gurevich2008dkal}\@. 

DKAL may be viewed as a conservative extension of disjunction-free intuitionistic logic~\cite{gurevich2009dkal},
where trust is definable. For example statements such as {\em 'k is trusted on saying x'}, for an infon $x$, logically means $(k\,said\,x) \Imp x$ in DKAL\@. 
This is reminiscent of the encoding of authority $(\Attest{K}{\phi}) \Imp \phi$ of principal $K$ over 
certain properties $\phi$ in Cyberlogic.
Both Cyberlogic and DKAL  are based on different extensions of intuitionistic logics with an attestion modality.
But DKAL chooses a disjunction-free propositional intuitionistic logic for its emphasis on efficient query answering, whereas
Cyberlogic is based on a more expressive first-order intuitionistic logic.
Thus Cyberlogic is more general than DKAL, at the expense of decidability and low complexity query answering. 
Moreover, and in contrast to Cyberlogic, query answering in DKAL is not based on executing distributed logic programs and it
does not support evidential authorization, even though there has been an initial proposal for an evidential DKAL, where communicated information is accompanied with sufficient justification~\cite{blass2011evidential}\@. 
Finally both distributed authorization logics distinguish between knowing and attesting, but the knowledge modality in Cyberlogic also supports knowledge shared by multiple principals.
Therefore it should be possible to embed DKAL-like authorization systems into Cyberlogic, and to benefit
from DKAL's efficient query answering for certain subclasses of authorization queries.

Verifiable computation based on probabilistically checkable proofs~\cite{arora1998proof} 
or fully homomorphic encryption~\cite{gentry2009fully}
lacks necessary features like public verification and their performance is currently not acceptable. 
In contrast, Pinocchio is a concrete system for efficient public verification of general computations with constant-sized proofs~\cite{parno2013pinocchio}\@. 
The applicability of non-interactive zero-knowledge proof methods (see, for example, \cite{mouris2021zilch})
to scalable and verifiable evidential transactions in Cyberlogic 
and concurrent zero-knowledge \cite{dwork2004concurrent} needs 
to be investigated.

Finally, a different proposal for cryptographic proofs from the one based on functional encryption in Section ~\ref{sec:crypto.proofs} implements a zero-knowledge proof of the fact that the proof in the hand of the prover/authorizer is a 
correct proof of the property in question~\cite{kumar2012proofs}\@.
Hereby, the property to be proven is seen as an encryption key while the proof term is the corresponding secret key.

\chapter{Conclusions}  \label{sec:conclusions}

Cyberlogic formulates a uniform semantics and logic for evidential transactions in open-ended, large-scale, and distributed environments. 
Its main concepts form the basis of a number of applications in digital government, digital forensics, digital rights management, access control, accountability, pervasive compliance, blockchain-based smart contracts, and software certification.
It is less clear, however, how one designs these distributed policies and protocols systematically and correctly, so that they can be shown to meet stated requirements, including all-important security and privacy requirements. 

Cyberlogic extends (first-order) intuitionistic logic with an attestation modality for specifying a rich variety of services, policies, protocols, and workflows. 
In this way, a number of essential features of authorization policies such as trust, revocation, and delegation can be encoded into Cyberlogic.
Trust, for instance, is definable by the authorization $\Attest{k}{\phi} \Imp \phi$, that is, we trust $\phi$ to hold whenever principal $k$ attests to it.
Moreover, temporal logic operators are definable in Cyberlogic based on a trusted time source. Cyberlogic with its (joint) knowledge modality and temporal logic operators therefore forms an expressive temporal-epistemic 
logic~\cite{engelfriet1996minimal} for reasoning about distributed policies and protocols. 
Another distinguishing feature is that Cyberlogic encodings are executable, and distributed proof search is used to construct evidence for authorizations and capabilities, which can be verified independently. 
Evidence in Cyberlogic relies on its realizability semantics and on signing direct attestations using public/private key pairs as associated with principals. 
The construction and communication of explicit proof terms, however, usually does not scale to large-scale distributed applications.
In addressing these efficiency issues we are currently experimenting with {\em certificate transparency}\footnote{
{\tt https://sites.google.com/site/certificatetransparency/log-proofs-work/}
} 
mechanisms~\cite{schoepp2021}, and the current implementation of ETB Datalog\footnote{{\tt https://git.fortiss.org/evidentia/etb}} supports Cyberlogic-like attestations based on IPFS.\footnote{{\tt https.io}}

Other requirements for a Cyberlogic interpreter include handling of failures of the underlying communication and computation substrate,
a specific mechanism for service invocation that is richer than the interpretation of built-in predicates~\cite{cruanes2014semantics},
incremental maintenance of policies and evidence,
and modern cryptographic schemes for privacy protection.
There are solutions for each of these issues, and the challenge is to seamlessly integrate 
them in a coherent and efficient Cyberlogic interpreter.

Altogether, Cyberlogic is a foundational logic for making transactions in large-scale distributed information systems such as the Internet trustworthy based on verifiable evidence and a small kernel of trust.
Its intent is to maximize the possibility of accountability for violations, thereby encouraging compliance~\cite{weitzner2008information}\@. 
This, however, requires digital evidence to hold up to scrutiny in real-world 
investigations, and it also requires a fine balance between relevant transparency and privacy needs.

\bibliographystyle{alpha}
\bibliography{sample}

\newpage
\begin{appendix}
\chapter{First Incarnation}
The developments in this report are based on the following invited presentation for introducing Cyberlogic at the National Security Agency’s third {\em High Confidence Software and Systems} Conference in Baltimore, MD on April 2003. 
\includepdf[pages=-, nup = 2x2]{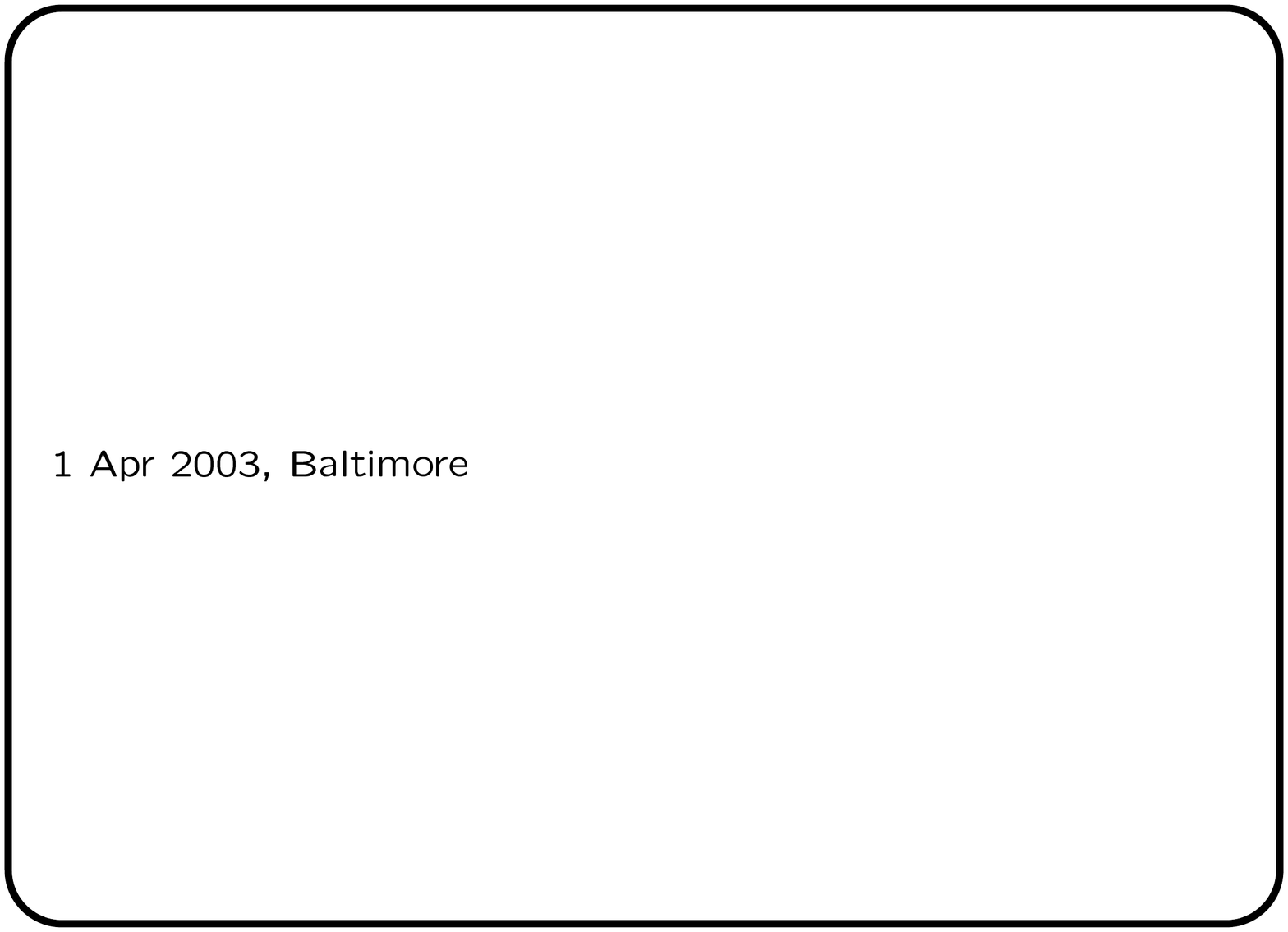}
\end{appendix}

\end{document}